\documentstyle[prd,aps]{revtex}
\def\be{\begin{equation}}
\def\ee{\end{equation}}
\def\ba{\begin{array}}
\def\ea{\end{array}}
\def\bea{\begin{eqnarray}}
\def\eea{\end{eqnarray}}

\begin{document}
\title{Octet and Decuplet Baryon Magnetic Moments in the Chiral Quark Model}
\author{Harleen Dahiya  and Manmohan Gupta}
\address {Department of Physics, Centre of Advanced Study in Physics,
Panjab University, Chandigarh-160 014, India.} 
\date{\today}
 \maketitle
\begin{abstract}
Octet and decuplet baryon magnetic moments 
have been formulated within
the chiral quark model ($\chi$QM)  with configuration mixing 
incorporating the sea quark  polarizations and
their orbital angular momentum through the generalization of the Cheng-Li 
mechanism.
When the parameters of $\chi$QM without configuration mixing
 are fixed by incorporating the
latest data pertaining to $\bar u-\bar d$ asymmetry (E866) and the spin 
polarization functions, in the case of octet magnetic moments 
the results not only show improvement 
over the nonrelativistic quark model results but also give a non zero 
value for  the right hand side of Coleman-Glashow sum rule, usually
zero in most of the models. In the case of decuplet magnetic
moments, we obtain a good overlap for $\Delta^{++}$, $\Omega^-$ 
and the  transition magnetic moment $\Delta N$ for which data 
are available. In the case of octet, the predictions of the  
$\chi$QM with the generalized Cheng-Li mechanism show remarkable 
improvements in general when effects of configuration mixing and 
``mass adjustments'' due to confinement are included, specifically 
in the case of $p$, $\Sigma^+$, $\Xi^o$, the $\Sigma \Lambda$ transition 
magnetic moment and in the violation of Coleman Glashow sum rule 
an almost perfect agreement with data is obtained.
When the above analysis is repeated with the earlier NMC data, a similar
level of agreement is obtained, however the results in the
case of E866 look to be better.
In case, we incorporate in our analysis the gluon polarization $\Delta g$,
found phenomenologically through the relation
$\Delta\Sigma(Q^2)=\Delta\Sigma-\frac{3\alpha_s(Q^2)}{2\pi}\Delta g(Q^2)$,
 we not only obtain improvement in the quark spin distribution functions 
and magnetic moments but also the value of $\Delta g$ comes out to 
be in good agreement with certain recent measurements as well as 
theoretical estimates.
\end{abstract}

\section{Introduction}

The measurements of the polarized structure functions of proton 
in the deep inelastic scattering (DIS) experiments
\cite{EMC,adams,abe} have shown that the valence quarks of the 
proton carry only about 30\% of its spin.
This ``unexpected'' conclusion from the point of view of 
nonrelativistic quark model (NRQM) becomes all the more intriguing
when it is realized that the NRQM is able to give a reasonably
good description of baryon octet magnetic moments using the assumption
that magnetic moments of quarks are proportional to the spin carried
by them. Further, this issue regarding spin and magnetic moments becomes 
all the more difficult to
understand when it is realized that the magnetic moments of
baryons receive contributions not only from the magnetic moments carried
by the valence quarks but also from various complicated effects,
such as orbital excitations \cite{{orbitex}}, 
relativistic and exchange current effects \cite{{mgupta1},{excurr}}, 
pion cloud contributions \cite{{pioncloudy}},  
effect of the confinement on quark masses \cite{{effm1},{effm2}}, 
effects of configuration mixing \cite{{mgupta1},{effm2},{Isgur}}, 
``quark sea'' polarizations \cite{{cheng},{chengsu3},{cheng1},{song},{johan}}, 
pion loop corrections \cite{loop}, etc.. 
Recently, it has been emphasized \cite{{johan},{cg1}} that the problem
regarding magnetic moments gets further complicated when one realizes
that the Coleman Glashow sum rule (CGSR) for octet magnetic moments \cite{cg},
valid in a large variety of models, is convincingly violated
by the data \cite{PDG}.

Recently, in a very interesting work, Cheng and Li  \cite{cheng1}
have shown that the DIS conclusions regarding the proton spin and the 
success of the NRQM in explaining magnetic moments can be reconciled in 
the $\chi$QM \cite{cheng,wein,manohar,eichten} 
if the $q \bar q$ sea, produced by the chiral fluctuations,
besides being polarized is also endowed with angular momentum.
In particular, in the case of the nucleon they have shown that the above 
mentioned mechanism (to be referred to as Cheng-Li mechanism) leads to
almost cancellations of the magnetic moment contribution of the 
polarized ``quark sea'' and its angular momentum leaving the description
of magnetic moment of nucleon in terms of the polarization of the 
valence quarks.
The authors, in a very recent Rapid Communication \cite{hdorbit},
by considering the generalization of the Cheng-Li mechanism to 
hyperons incorporating coupling breaking and mass breaking terms, found that
one is able to get a
non zero value for the violation of CGSR ($\Delta$CG) \cite{CGSR}  
apart from       
improving the NRQM predictions for magnetic moments of the octet baryons. 
This fact, when viewed in the context of success of 
$\chi$QM  \cite{cheng,chengsu3,cheng1,song,johan,eichten}, 
for the explanation of $\bar u-\bar d$
asymmetry \cite{GSR,E866,NMC}, existence of significant
strange quark content \cite{{EMC},{adams},{abe}},
quark flavor and spin distribution functions \cite{adams}, 
hyperon decay parameters etc., 
strongly indicates that constituent quarks,  
weakly interacting Goldstone bosons (GBs) and $q \bar q$ pairs
provide the appropriate degrees of freedom at the leading order
in the scale between chiral
symmetry breaking ($\chi_{SB}$) and the  confinement scale.
This is further borne out by the fact that when the generalized Cheng-Li
mechanism is combined with the effects of configuration mixing, known to
be improving the predictions of NRQM
\cite{{mgupta1},{Isgur},{DGG},{yaouanc},{photo}} as well as
compatible \cite{{riska},{chengspin},{prl}} with $\chi$QM, and
``mass adjustments'' arising due to confinement of quarks  
\cite{{effm1},{effm2}}, leads to an
almost perfect fit for $\Delta$CG and an excellent fit for octet
magnetic moments \cite{hdorbit}.
In view of this, it is desirable to
broaden the scope of Ref \cite{hdorbit} by extending the calculations to 
decuplet magnetic moments, transition magnetic moments and by delving
into the detailed implications of some of the crucial ingredients 
such as the generalized Cheng-Li mechanism
(with and without configuration mixing) and ``mass adjustments'' on the octet
magnetic moments, not detailed in Ref \cite{hdorbit}.
At the same time, for an appropriate appraisal of the implications 
of the calculated magnetic
moments, it is desirable to fine tune the $\chi$QM parameters by analysing 
the latest data pertaining to $\bar u-\bar d$ asymmetry \cite{E866}, 
spin polarization functions \cite{adams} as well as the flavor 
non-singlet components.

The purpose of the present paper is to detail the formulation of the octet and
decuplet magnetic moments in $\chi$QM incorporating the
generalized Cheng-Li mechanism (with and without configuration mixing). 
In order to make our analysis regarding magnetic moments more
responsive, we have carried out a brief analysis to fix the 
$\chi$QM parameters using the latest data 
regarding the quark distribution functions and spin distribution functions. 
A brief discussion on the flavor singlet component of the total helicity
including gluon polarization
and its implications on the magnetic moments is also very much in order.
Further, we also intend to study the
implications of variation of quark masses  as well as the angle 
pertaining to configuration mixing on magnetic moments.

The plan of the paper is as follows. To make the manuscript readable
as well as to facilitate discussion, in Sec \ref{detail} we present
some of the essentials of $\chi$QM and Cheng-Li mechanism with an
emphasis on the details of its generalization. 
In Sec \ref{spin},  the modifications due to configuration mixing 
on the generalized Cheng-Li mechanism have been discussed. 
Sec \ref{inputs} includes a discussion on the various inputs
used in the analysis, in particular  the $\chi$QM parameters have
been obtained by fitting $\chi$QM with and without configuration mixing
to the latest data.
In Sec \ref{results}, we present the numerical results
and their discussion including a brief reference to the flavor singlet 
component as well as gluon polarization. 
Sec \ref{summary} comprises the
summary and the conclusions. 
To make the manuscript self contained, in the Appendix a few typical cases
pertaining to octet as well as decuplet baryons have been fully worked 
out.

\section{Magnetic moments in the $\chi$QM with the 
generalized Cheng-Li mechanism}
 \label{detail}

The basic process in the $\chi$QM is the
emission of a GB by a constituent quark which further splits into a $q
\bar q$ pair, for example,                          

\be
  q_{\pm} \rightarrow {\rm GB}^{0}
  + q^{'}_{\mp} \rightarrow  (q \bar q^{'})
  +q_{\mp}^{'}\,,                              \label{basic}
\ee
where $q \bar q^{'}  +q^{'}$ constitute the ``quark sea''
  \cite{{chengsu3},{cheng1},{song},{johan}}. 
The effective Lagrangian describing interaction between quarks and a nonet of
GBs, consisting of octet and a singlet, can be expressed as

\be
{\cal L} = g_8 \bar q \Phi q\,,
\ee
\bea 
q =\left( \ba{c} u \\ d \\ s \ea \right),&
~~~~~& \Phi = \left( \ba{ccc} \frac{\pi^o}{\sqrt 2}
+\beta\frac{\eta}{\sqrt 6}+\zeta\frac{\eta^{'}}{\sqrt 3} & \pi^+
  & \alpha K^+   \\
\pi^- & -\frac{\pi^o}{\sqrt 2} +\beta \frac{\eta}{\sqrt 6}
+\zeta\frac{\eta^{'}}{\sqrt 3}  &  \alpha K^o  \\
 \alpha K^-  &  \alpha \bar{K}^o  &  -\beta \frac{2\eta}{\sqrt 6}
 +\zeta\frac{\eta^{'}}{\sqrt 3} \ea \right), \eea
where $\zeta=g_1/g_8$, $g_1$ and $g_8$ are the coupling constants for the 
singlet and octet GBs, respectively.

SU(3) symmetry breaking is introduced by considering
$M_s > M_{u,d}$ as well as by considering
the masses of GBs to be nondegenerate
 $(M_{K,\eta} > M_{\pi})$ {\cite{{chengsu3},{cheng1},{song},{johan}}}, whereas 
  the axial U(1) breaking is introduced by $M_{\eta^{'}} > M_{K,\eta}$
{\cite{{cheng},{chengsu3},{cheng1},{song},{johan}}}.
The parameter $a(=|g_8|^2$) denotes the transition probability
of chiral fluctuation
of the splittings  $u(d) \rightarrow d(u) + \pi^{+(-)}$, whereas 
$\alpha^2 a$, $\beta^2 a$ and $\zeta^2 a$ respectively 
denote the probabilities of transitions of~
$u(d) \rightarrow s  + K^{-(o)}$, $u(d,s) \rightarrow u(d,s) + \eta$,
 and $u(d,s) \rightarrow u(d,s) + \eta^{'}$.

Following Cheng and Li \cite{cheng1}, the magnetic moment of a given baryon
which receives contributions from valence quarks, sea quarks and
the orbital angular momentum of the ``quark sea'' is expressed as
\be
\mu(B)_{{\rm total}} = \mu(B)_{{\rm val}} + \mu(B)_{{\rm sea}} +
\mu(B)_{{\rm orbit}}\,.     \label{totalmag}
\ee
The valence and the sea contributions, in terms of quark spin
polarizations, can be written as 
\be
\mu(B)_{{\rm val}}=\sum_{q=u,d,s} {\Delta q_{{\rm val}}\mu_q} ~~~ 
{\rm and } ~~~
\mu(B)_{{\rm sea}}=\sum_{q=u,d,s} {\Delta q_{{\rm sea}}\mu_q}\,,    \label{mag}
\ee
where $\mu_q= \frac{e_q}{2 M_q}$ ($q=u,d,s$) is the quark magnetic moment,
$e_q$ and $M_q$ are the electric charge and the mass respectively for the
quark $q$. Similarly, the orbital angular momentum 
contribution of the sea, $\mu(B)_{{\rm orbit}}$, can be expressed in
terms of the valence   
quark polarizations and the orbital moments of the sea quarks, 
the details of which would be given in Section \ref{seaorbit}. 
Following references {\cite{{cheng},{cheng1},{johan}}}, 
the quark spin polarization can be defined
as 
\be
\Delta q= q_{+}- q_{-}+
\bar q_{+}- \bar q_{-}\,,   \label{spin contr}
\ee
where $q_{\pm}$ and $\bar q_{\pm}$ can be 
calculated from the spin
structure of a baryon defined as 
\be
\hat B \equiv \langle B|\cal N|B \rangle\,, \label{BNB}
\ee 
where $|B\rangle$ is the baryon wave function  and $\cal N$ is the number 
operator, for example,
\be
{\cal N}=n_{u_{+}}u_{+} + n_{u_{-}}u_{-} + n_{d_{+}}d_{+} + n_{d_{-}}d_{-} +
n_{s_{+}}s_{+} + n_{s_{-}}s_{-}\,, \label{number}
\ee  
with the coefficients of the $q_{\pm}$ giving the number of 
$q_{\pm}$ quarks.

To calculate $\mu(B)_{{\rm val}}$, we need to calculate the valence
spin polarizations $\Delta q_{{\rm val}}$. 
For ready reference some essential details of the calculations for
valence quark polarizations pertaining to typical cases are presented 
in Appendix A. 

\subsection{ Contribution of the ``quark sea'' polarizations to
the magnetic moments}
To evaluate the ``quark sea'' magnetic moment, 
one has to find ${\Delta q_{{\rm sea}}}$ corresponding to each baryon. 
For detailed evaluation of  ${\Delta q_{{\rm sea}}}$, we refer the 
reader to Refs. \cite{{chengsu3},{cheng1},{song},{johan}}, however to 
facilitate its extension to the case with configuration mixing, we 
summarize some of the essentials adopted for the present use.
The spin structure for the process given in Eq. (\ref{basic}), 
after one interaction, can be obtained  by  substituting for every
valence quark, for example,  

\be
 q_{\pm} \rightarrow \sum P_q q_{\pm} +
 |\psi(q_{\pm})|^2\,, \label{q}
\ee
where $\sum P_q$ is the probability of emission of GB from a $q$ quark and the 
probabilities of transforming a $q_{\pm}$ quark are
$|\psi(q_{\pm})|^2$. The relevant details pertaining to the
 calculations of ${\Delta q_{{\rm sea}}}$, again for some typical cases, 
are presented in  Appendix A. The expressions for ${\Delta
 q_{{\rm sea}}}$ in the case of proton are as follows
\be
\Delta u_{\rm sea}=-\frac{a}{3} (7+4 \alpha^2+
 \frac{4}{3}\beta^2 +\frac{8}{3} \zeta^2)\,, ~~~
\Delta d_{\rm sea}=-\frac{a}{3} (2-\alpha^2
-\frac{1}{3}\beta^2 -\frac{2}{3} \zeta^2)\,,~~~
 \Delta s_{\rm sea}=-a \alpha^2\,.
\ee
The expressions for the other octet baryons can be found from
Table \ref{ocsea}.

The  ``quark sea'' spin polarizations for the decuplet baryons
can be calculated  in a similar manner as that of octet baryons. For
example, the general expressions for the spin structure of the decuplet 
baryons of the types $B^*(xxy)$, $B^*(xxx)$ and $B^*(xyz)$, using
Eq. (\ref{q}), are respectively given as
\be
{\hat B}^*(xxy)= 2 \left(\sum P_x x_{+} +{|\psi(x_{+})|}^2\right)
+ \left(\sum P_y y_{+}+ {|\psi(y_{+})|}^2\right),
\ee
\be
{\hat B}^*(xxx)= 3 \left(\sum P_x x_{+} + {|\psi(x_{+})|}^2\right),
\ee
\be
{\hat B}^*(xyz)= \left(\sum P_x x_{+} + {|\psi(x_{+})|}^2\right)+
\left(\sum P_y y_{+} + {|\psi(y_{+})|}^2\right)
+ \left(\sum P_z z_{+} + {|\psi(z_{+})|}^2\right),
\ee
where $x$, $y$ and $z$ correspond to any of the $u$, $d$ and $s$ quarks.
The detailed expressions for the spin polarizations $\Delta q_{{\rm sea}}$, 
corresponding to the decuplet baryons, can again be found from 
Table \ref{ocsea}.

\subsection{Contribution of the ``quark sea'' orbital angular momentum to
the magnetic moments} \label{seaorbit} 

Following  Cheng and Li \cite{cheng1}, the contribution of the 
angular momentum of the ``quark sea'' to the magnetic moment of a given 
quark is given as 
\be
\mu (q_{+} \rightarrow {q}_{-}^{'}) =\frac{e_{q^{'}}}{2M_q}
\langle l_q \rangle +
\frac{{e}_{q}-{e}_{q^{'}}}{2 {M}_{{\rm GB}}}\langle {l}_{{\rm GB}} \rangle\,,
\ee
where 
\be
\langle l_q \rangle=\frac{{M}_{{\rm GB}}}{M_q+{M}_{{\rm GB}}} ~{\rm and} 
~\langle l_{{\rm GB}} \rangle=\frac{M_q}{M_q+{M}_{{\rm GB}}}\,,
\ee
$\langle l_q, l_{{\rm GB}} \rangle$ and ($M_q$, ${M}_{{\rm GB}}$) are the 
orbital angular momenta and masses of quark and GB respectively.
The orbital moment of each process is then multiplied by the probability
for such a process to take place to yield the magnetic moment due to
all the transitions starting with a given valence quark, for example,
\be [ \mu (u_{\pm}(d_{\pm}) \rightarrow )]=  \pm a
\left [\mu (u_{+}(d_{+}) \rightarrow d_- (u_-)) +
\alpha^2 \mu (u_+(d_+) \rightarrow s_-) 
+(\frac{1}{2} +\frac{1}{6} \beta^2+ \frac{1}{3} \zeta^2)
\mu (u_{+}(d_{+}) \rightarrow u_- (d_-))\right ], \label{mud}
\ee

\be
[\mu (s_{\pm} \rightarrow )]=  \pm a
\left [\alpha^2 \mu (s_{+} \rightarrow u_-) +
\alpha^2 \mu (s_+ \rightarrow d_-) +
(\frac{2}{3} \beta^2+ \frac{1}{3} \zeta^2)
\mu (s_{+} \rightarrow s_- ) \right ]. \label{mus}
\ee
The above equations, derived by Cheng and Li, alongwith 
$\Delta q_{\rm sea}$ constitute the essentials of Cheng-Li mechanism. 
Eqs. (\ref{mud}) and (\ref{mus}) can easily be
generalized by including the coupling breaking and mass
breaking terms. For example, in terms of the parameters $a$, $\alpha$,
$\beta$ and $\zeta$, the  orbital moments of $u$, $d$ and $s$ quarks 
respectively are
\bea
\mu(u_+ \rightarrow) & =& 
a \left [\frac{-M^2_{\pi}+3 M^2_{u}}{2 {M}_{\pi}(M_u+{M}_{\pi})}
-\frac{\alpha^2(M^2_{K}-3 M^2_{u})}{2 {M}_{K}(M_u+{M}_{K})} 
+ \frac{(3+\beta^2+2 \zeta^2)M^2_{\eta}}{6 {M}_{\eta}(M_u+{M}_{\eta})}
 \right ]{\mu}_N\,, \label{orbitu} \\ 
\mu(d_+ \rightarrow) & =& 
a \frac{M_u}{M_d}\left [\frac{2 M^2_{\pi}-3 M^2_{d}}
{2 {M}_{\pi}(M_d+{M}_{\pi})}-
 \frac{\alpha^2 M^2_{K} }{2 {M}_{K}(M_d+{M}_{K})}   
- \frac{(3+\beta^2+2 \zeta^2)M^2_{\eta}}{12 {M}_{\eta}(M_d+{M}_{\eta})}
 \right ]{\mu}_N\,, \label{orbitd}  \\  
\mu(s_+ \rightarrow) & =& 
a \frac{M_u}{M_s}\left [ \frac{\alpha^2 (M^2_{K}-3 M^2_s) }
{2 {M}_{K}(M_s+{M}_{K})}  -
\frac{(2 \beta^2+\zeta^2)M^2_{\eta}}{6 {M}_{\eta}(M_s+{M}_{\eta})}
 \right ]{\mu}_N\,, \label{orbits}
\eea
where $\mu_N$ is the nuclear magneton. Eqs. (\ref{orbitu}),
(\ref{orbitd}) and (\ref{orbits}) alongwith  $\Delta q_{\rm sea}$
shall be referred to as the generalized Cheng-Li mechanism.  
The orbital contribution to the magnetic moment of the octet baryon 
of the type $B(xxy)$ in terms of the above equations as well as the valence 
spin polarizations is given as
\be
\mu(B)_{{\rm orbit}} =\Delta x_{{\rm val}} \left[\mu (x_+ \rightarrow) \right]+
\Delta y_{{\rm val}} \left[\mu (y_+ \rightarrow) \right].  \label{orbit}
\ee

Similarly, the orbital contributions in the case of the decuplet baryons 
$B^*(xxy)$,  $B^*(xxx)$ and $B^*(xyz)$ are respectively given as
\be
\mu(B^*)_{{\rm orbit}} =\Delta x_{{\rm val}} [\mu (x_+ \rightarrow)] + 
\Delta y_{{\rm val}}[\mu (y_+ \rightarrow)]\,,
\label{de orbit}
\ee
\be
\mu(B^*)_{{\rm orbit}} = \Delta x_{{\rm val}} [\mu (x_+ \rightarrow)]\,,
\ee
\be
\mu(B^*)_{{\rm orbit}} =  \Delta x_{{\rm val}} [\mu (x_+ \rightarrow)] + 
\Delta y_{{\rm val}}[\mu (y_+ \rightarrow)]
+\Delta z_{{\rm val}} [\mu (z_+ \rightarrow)]\,.
\ee

\section {Generalized Cheng-Li mechanism with configuration mixing} 
\label{spin}

Spin-spin forces,  known to be compatible 
\cite{{riska},{chengspin},{prl}} with the $\chi$QM, generate
configuration mixing \cite{{Isgur},{DGG},{yaouanc}} for the
octet baryons which effectively leads to modification
of the valence quark and ``quark sea'' spin distribution functions \cite{hd}.
From Eqs. (\ref{mag}) and (\ref{orbit}), it is evident that
the effects of configuration mixing on magnetic moments
can be included if one is able to
estimate the same on the valence and sea contributions to magnetic moments.
The most general configuration mixing generated by the spin-spin forces
in the case of octet baryons \cite{{Isgur},{yaouanc},{full}} 
can be expressed as
\be
|B \rangle=\left(|56,0^+\rangle_{N=0} \cos \theta +|56,0^+ \rangle_{N=2}  
\sin \theta \right) \cos \phi 
+  \left(|70,0^+\rangle_{N=2} \cos \theta^{'} +|70,2^+\rangle_{N=2}  
\sin \theta^{'} \right) \sin \phi\,, \label{full mixing}
\ee
where $\phi$ represents the $|56\rangle-|70\rangle$ mixing whereas
$\theta$ and  $\theta^{'}$ respectively correspond to the mixing 
among $|56,0^+\rangle_{N=0}-|56,0^+ \rangle_{N=2}$ states and
$|70,0^+\rangle_{N=2}-|70,2^+\rangle_{N=2}$ states.
For the present purpose, it is adequate
{\cite{{mgupta1},{effm2},{yaouanc},{hd}}} to consider the mixing only between
$|56,0^+ \rangle_{N=0}$ and the $|70,0^+\rangle_{N=2}$ states, for
example,

\begin{equation}
|B\rangle 
\equiv \left|8,{\frac{1}{2}}^+ \right> 
= \cos \phi |56,0^+\rangle_{N=0}
+ \sin \phi|70,0^+\rangle_{N=2}\,,  \label{mixed}
\end{equation} 
where
\bea
 |56,0^+\rangle_{N=0} &=& \frac{1}{\sqrt 2}(\chi^{'} \phi^{'} +
\chi^{''} \phi^{''}) \psi^{s}(0^+)\,, \label{56}   \\      
|70,0^+\rangle_{N=2} &=&  \frac{1}{2}[(\phi^{'} \chi^{''} +\phi^{''}\chi^{'})
\psi^{'}(0^+) + (\phi^{'} \chi^{'} -\phi^{''} \chi^{''})\psi^{''}(0^+)]\,, 
\label{70}
\eea
with
\be
\chi^{'} =  \frac{1}{\sqrt 2}(\uparrow \downarrow \uparrow
-\downarrow \uparrow \uparrow)\,,~~~
\chi^{''} =  \frac{1}{\sqrt 6} (2\uparrow \uparrow \downarrow
-\uparrow \downarrow \uparrow -\downarrow \uparrow \uparrow)\,, 
\ee 
representing the spin wave functions.
In general, the isospin wave functions for the octet baryons of the 
type $B(xxy)$ are   given as
\be
\phi^{'}_B = \frac{1}{\sqrt 2}(xyx-yxx)\,,~~~
\phi^{''}_B = \frac{1}{\sqrt 6}(2xxy-xyx-yxx)\,,
\ee
whereas for $\Lambda (xyz)$  and  $\Sigma^o (xyz)$ they are  given as
\bea
 \phi^{'}_{\Lambda} &=& \frac{1}{2 \sqrt 3}(xzy+zyx-zxy-yzx-2xyz-2yxz)\,,~~~
\phi^{''}_{\Lambda} = \frac{1}{2}(zxy+xzy-zyx-yzx)\,, \\
 \phi^{'}_{\Sigma^o}&=& \frac{1}{2}(zxy+zyx-xzy-yzx)\,,~~~
\phi^{''}_{\Sigma^o}=\frac{1}{2 \sqrt 3}(zyx+zxy+xzy+yzx-2xyz-2yxz)\,.
\eea
 For the definition of the spatial  wave functions
 ($\psi^{s}, \psi^{'}, \psi^{''})$ as well as the
 definitions of the overlap integrals, we  refer the
 reader to reference {\cite{{yaoubook}}.
The mixing expressed through the Eq. (\ref{mixed}) would be referred to as 
the ``mixed'' octet, henceforth we will not distinguish between
configuration mixing and the ``mixed'' octet.

Using the above wavefunctions, one can easily find the spin polarizations for 
proton, for example
\be
\Delta u_{{\rm val}} ={\cos}^2 \phi \left[\frac{4}{3} \right] 
   + {\sin}^2 \phi \left[\frac{2}{3}  \right],~~~
\Delta d_{{\rm val}} ={\cos}^2 \phi \left[-\frac{1}{3} \right]  +
  {\sin}^2 \phi \left[\frac{1}{3}  \right],~~~
\Delta s_{{\rm val}} = 0\,. \label{sigma1}
\ee 
These expressions would replace $\Delta q_{{\rm val}}$
in Eq. (\ref{mag}) and (\ref{orbit}) for  calculating the effects of 
configuration mixing on the valence and the orbital part in the case
of proton. Similarly, one can easily find the spin polarization functions
for other ``mixed'' octet members.

The ``quark sea'' polarizations also gets modified with the
inclusion of configuration mixing and 
can easily be calculated,  the details of the calculations in the case of
$p$, $\Lambda$ and $\Sigma \Lambda$ are given in Appendix A. 
For the case of proton, these are expressed as
\bea
\Delta u_{\rm sea}&=&-{\cos}^2 \phi \left[\frac{a}{3} (7+4 \alpha^2+
 \frac{4}{3}\beta^2 +\frac{8}{3} \zeta^2)\right]
-{\sin}^2 \phi \left[\frac{a}{3} (5+2 \alpha^2
+\frac{2}{3}\beta^2 +\frac{4}{3} \zeta^2)\right], \\
\Delta d_{\rm sea}&=&-{\cos}^2 \phi \left[\frac{a}{3} (2-\alpha^2
-\frac{1}{3}\beta^2 -\frac{2}{3} \zeta^2)\right]
-{\sin}^2 \phi \left[\frac{a}{3} (4+\alpha^2
+\frac{1}{3}\beta^2 +\frac{2}{3} \zeta^2)\right], \\ 
 \Delta s_{\rm sea}&=&-a \alpha^2. \label{sigma2}
\eea 
The ``quark sea'' spin polarizations for the other octet baryons and
transition magnetic moments can
similarly be calculated and are presented in 
Table \ref{ocsea}. 

Configuration mixing due to spin-spin forces does not affect decuplet 
baryons \cite{Isgur,yaouanc}, 
thus the decuplet baryon wave function is given as 
 \be
|B^*\rangle \equiv |56,0^+\rangle_{N=0} =\chi^{s}\phi^{s} \psi^{s}(0^+)\,,
\label{56decuplet}
\ee
with
\be
\chi^{s} =  (\uparrow \uparrow \uparrow)\,. 
\ee
The isospin wave functions for the decuplet baryons of the types
$B^*(xxx)$, $B^*(xxy)$ and $B^*(xyz)$ respectively are
\be
\phi^{s}_{B^*} = xxx\,, ~~~
\phi^{s}_{B^*} = \frac{1}{\sqrt 3}(xxy + xyx + yxx)\,,~~~
\phi^{s}_{B^*} = \frac{1}{\sqrt 6}(xyz + xzy + yxz + yzx + zxy + zyx)\,, 
\ee
where $x$, $y$ and $z$ correspond to any of the $u$, $d$ and $s$ quarks.

\section {Inputs} \label{inputs}
To facilitate the understanding of different inputs based on 
Eq. (\ref{totalmag}), in Appendix A we have
presented the complete expressions for two of the octet
baryon magnetic moments $p$ and $\Lambda$ as well as
the $\Sigma \Lambda$ transition magnetic moment, for the
case of decuplet baryons we have considered the example of $\Delta^+$. 
The other octet and decuplet magnetic moments can
similarly be formulated. As is evident from the Appendix, to calculate
magnetic moments we need inputs related to $\chi$QM parameters, 
mixing angle $\phi$  and quark masses. 
The parameters $a$, $\alpha$,
$\beta$ and $\zeta$ of the $\chi$QM, are usually fixed by considering
the spin polarization functions $\Delta u$, $\Delta d$, $\Delta s$
\cite{adams} and related $Q^2$ independent 
parameters $\Delta_3=\Delta u-\Delta d$ and 
$\Delta_8=\Delta u+ \Delta d-2 \Delta s$ \cite{ellis} as well as  
the  quark distribution functions including the violation of Gottfried sum
rule  \cite{{E866},{NMC}} measured through the  $\bar u-\bar d$ asymmetry.
In the present analysis we
have taken the  pion fluctuation parameter $a$ to be 0.1, 
in accordance with most of the other
calculations \cite{{cheng1},{song},{johan}}.
It has been shown \cite{johan} that to fix the
violation of Gottfried sum rule \cite{GSR}, we have to  consider the
relation
\be 
\bar u-\bar d=\frac{a}{3}(2 \zeta+\beta-3)\,.
\ee
In this relation, one immediately finds that in case the value of $a$
is taken to be 0.1 then to reproduce $\bar u-\bar d$ asymmetry one
gets the relation $\zeta=-0.3-\beta/2$ for the E866 data \cite{E866} 
and $\zeta=-0.7-\beta/2$ for the case of NMC data \cite{NMC}.
Before carrying out the analysis of $\chi$QM with configuration mixing
one has to fix the mixing angle $\phi$ which in the present case is taken
to be $\phi=20^o$ by fitting the neutron charge radius 
\cite{{yaouanc},{full},{neu charge}}. After carrying out our analysis
regarding the spin polarization functions and using the 
latest E866 \cite{E866} data and the NMC \cite{NMC} data regarding the
$\bar u-\bar d$ asymmetry, in Table \ref{ud}, we have presented the 
calculated values of 
certain phenomenological quantities having implications for 
$\chi$QM parameters ($\alpha$ and $\beta$) with and without
configuration mixing. 
From the table we find that chiral quark model with configuration mixing
($\chi$QM$_{gcm}$) is able to give a fairly
good fit to the various spin distribution functions as well 
as quark distribution functions, in particular, the agreement in the
case of $\Delta_3$, $\Delta_8$, $f_s$, $f_3/f_8$ is quite striking.
In the table we have not included the flavor singlet component of 
the total helicity ($\Delta \Sigma =\Delta u+\Delta d+ \Delta s$) which 
shall be discussed later separately. The $\chi$QM parameters thus found 
are summarized in Table \ref{input} and constitute the input for
magnetic moment calculations.

The orbital angular moment contributions  are  characterized by
the parameters of $\chi$QM as well as the masses of the GBs. 
For evaluating the contribution  of pions, 
we have used its on mass shell value in accordance with
several other similar calculations \cite{{mpi1}}. Similarly, for the
other GBs we have considered their on mass shell values, however their
contributions are much smaller compared to the pionic contributions.

In accordance with the basic assumptions of $\chi$QM, the constituent
quarks are supposed to have only Dirac magnetic moments governed by
the respective quark masses.
In the absence of any definite guidelines for the constituent quark
masses, for the $u$ and $d$ quarks we have used  their most widely 
accepted values in hadron spectroscopy 
\cite{{Isgur},{chengspin},{yaoubook},{close},{mu1}}, 
for example $M_u=M_d=330$ MeV. Apart from taking the above quark masses, 
one has to consider the strange quark
mass implied by the various sum rules derived from the spin-spin
interactions for different baryons {\cite{{mgupta1},{Isgur},{yaouanc}}}, 
for example,
$ \Lambda-N =M_s-M_u$, $(\Sigma^*-\Sigma)/(\Delta-N)=M_u/M_s$ and 
$(\Xi^*-\Xi)/(\Delta-N)=M_u/M_s$, respectively fix 
$M_s$ for $\Lambda$, $\Sigma$ and $\Xi$ baryons. 
These quark masses and corresponding magnetic moments
have to be further adjusted by the quark confinement effects
\cite{{effm1},{effm2}}. In conformity with  additivity assumption, 
the simplest way to incorporate this adjustment \cite{{effm1},{effm2}} 
is to first express $M_q$ in the magnetic moment
operator in terms of $M_B$, the mass of the baryon obtained
additively from the quark masses, which then
is replaced  by $M_B+\Delta M$,
$\Delta M$
being the mass difference between the experimental value and $M_B$.
This leads to the following adjustments in the quark
magnetic moments:
$\mu_d = -[1-(\Delta M/M_B)] {\mu}_N$,
$\mu_s = -M_u/M_s [1-(\Delta M/M_B)]{\mu}_N$ and $\mu_u=-2 \mu_d$.
The baryon magnetic moments calculated after incorporating this effect
would be referred to as ``mass adjusted''. 

\section{Results and Discussions} \label{results}
Using Eq. (\ref{totalmag}) and the inputs discussed above as well as 
the expressions given in Table  \ref{ocsea}, in Table  \ref{E866} we have
presented the results of octet magnetic moments without taking any of
these as inputs.
For a general discussion of the contents of Table \ref{E866} we refer 
the readers to Ref \cite{hdorbit}, however, in the present case we would 
like to discuss in detail the role of generalized Cheng-Li
mechanism, configuration mixing and ``mass adjustments'' in getting the
fit for octet magnetic moments. To this end, one can immediately find that
$\chi$QM with generalized Cheng-Li mechanism,  however without 
configuration mixing
and ``mass adjustments'', consistently improves the predictions of NRQM
as well as is able to generate a non zero value of $\Delta$CG. 
On closer examination of the results, several interesting points pertaining
to generalized Cheng-Li mechanism emerge.
The total contribution
to the magnetic moment is coming from several sources with similar and
opposite signs, for example, the orbital is contributing with the same
sign as the valence part, whereas the sea is contributing with opposite
sign. The sea and orbital contributions are fairly 
significant as compared to the valence contributions 
and they cancel in the right direction, for example,
the valence contributions of $p$, $\Sigma^+$ and $\Xi^o$ are
 higher in magnitude than the experimental
value but the sea contribution being higher in magnitude than the orbital
contribution reduces the valence contribution leading to a 
better agreement with data. Similarly, in the case of $n$, $\Sigma^-$ 
and $\Sigma \Lambda$ the valence contribution in magnitude is lower 
than the experimental value but in these cases the sea contribution 
is lower than the orbital part so it adds on to the valence contribution 
again improving agreement with data. 
Thus, in a very interesting manner, the orbital and sea contributions
together add on to the valence contributions leading to better agreement
with data as compared to NRQM. This not only endorses the earlier
conclusion of Cheng and Li \cite{cheng1} but also suggests that the Cheng-Li
mechanism could perhaps provide the dominant dynamics of the
constituents in the nonperturbative regime of QCD on
which further corrections could be evaluated.
To this end, in Table \ref{E866}, we have presented the
results wherein the effects of configuration mixing and ``mass adjustments''
have been included.
As is evident from the table, we have
been able to get an excellent fit for
almost all the baryons, it is almost perfect for $p$, $\Sigma^+$, $\Xi^o$,
$\Sigma \Lambda$ and  $\Delta$CG
whereas in the other cases the value is reproduced
within 5\% of data.

In order to study closely the role of configuration mixing on octet
magnetic moments, in Table  \ref{E866} we have presented the
results with and without mixing, however with the inclusion of 
``mass adjustments''. 
As is evident from the table, one finds that the individual 
magnetic moments show improvements after the inclusion of configuration 
mixing, particularly  in the case of $p$, $n$, $\Sigma^+$, $\Xi^o$,  
$\Lambda$ and $\Sigma \Lambda$ one observes a significant improvement. 
It may be noted that configuration mixing reduces
valence, sea and orbital contributions to the magnetic moments
and the results which are generally on the higher side get
corrected in the right direction by the inclusion of 
configuration mixing.
This is particularly manifest in the case of $\Xi$ particles, for
example, the magnitude of $\Xi^o$ magnetic moment without configuration 
mixing is lowered so as to achieve an almost perfect fit,
whereas in case of $\Xi^-$, a difficult case for most of the models,
configuration mixing increases the magnitude for better
agreement with data.  
In contrast to general improvement in the case of individual magnetic 
moments, $\Delta$CG hardly gets affected by configuration mixing.
In view of the fact that $\chi$QM with configuration  mixing involves
baryon wave functions which are perturbed by the spin-spin forces,
therefore, in principle one should employ the fully perturbed wave 
functions of the
octet baryons as derived by Isgur {\it et al.} \cite{Isgur} 
given in Eq. (\ref{full mixing}). 
However, we have found that for the present case the use of
``mixed'' octet (Eq. (\ref{mixed})) is adequate to reproduce the results
of fully perturbed wave function to the desired level of accuracy.
One may wonder 
whether $\Delta$CG could also be reproduced with the variation of 
mixing angle $\phi$. 
Our calculations in this regard show that variation 
of $\phi$ does not lead to any improvement in the magnetic moments as well as
$\Delta$CG. The present value of angle  $\phi$, fixed from the 
neutron charge radius  \cite{{yaouanc},{full},{neu charge}},
seems to be providing the best fit.

It would also perhaps be interesting to find out the 
implications of configuration mixing for $\chi$QM without ``mass adjustments''.
Broadly speaking the individual magnetic moments can again be fitted, 
however $\Delta$CG leaves much to be desired. This can be easily checked
from Table \ref{NMC}, wherein we have presented these calculations with 
the NMC data, the E866 based fit  follows the same 
pattern.
The value of $\Delta$CG registers a remarkable improvement
when effects due to ``mass adjustments'' alongwith  configuration mixing
are included. This is not 
surprising as the large value of $\Delta$CG could come only from the
valence quark corrections,  duly provided by the ``mass
adjustments''. It would be desirable to know what level of fit can be
achieved without  configuration mixing, however with the inclusion of 
``mass adjustments''.
A closer examination of the table immediately brings out that in this 
case the  individual magnetic moments leave much to be desired
whereas one is able to reproduce $\Delta$CG, in accordance with our 
earlier conclusions. It may also be noted that the ``mass
adjustments'' generally lower the various contributions
except for the nucleon.
In short, we may emphasize that the final fit obtained here
cannot be achieved if any of the ingredients, for example,
generalized Cheng-Li mechanism, configuration mixing and 
``mass adjustments, is absent.

For the sake of completeness, as mentioned earlier also, we have
presented in Table \ref{NMC} the octet magnetic moments when the
$\chi$QM parameters are 
fitted by incorporating NMC data. This table also includes our results
wherein magnetic moments have been calculated with configuration mixing
however without``mass adjustments'', not included in Table
\ref{E866}. From the table, one can immediately find out that the
basic pattern of results remain the same, however in general the
results are lower as compared to the case of E866 data. This is not
difficult to understand when one realizes that the contribution of sea
polarization in case of E866 and NMC data are quite different.
This can
be  understood easily when one realizes that the 
sea quark polarization is proportional to the  parameter $\zeta$.
Because of $|\zeta_{E866}|<|\zeta_{NMC}|$, one can easily
understand the corresponding lowering of the magnetic moments
in the case of NMC data, however both the calculations are in
good agreement with each other.

In Table \ref{decuplet}, we have presented the results of the decuplet 
baryons for the the latest E866  and the NMC data. 
The calculations of decuplet magnetic moments have been carried out
with the same $\chi$QM parameters and quark masses 
as that of the octet magnetic moments. From
the table, it is evident that we have been able to obtain a very 
good agreement pertaining to the case of $\Delta^{++}$ and $\Omega^-$ 
whereas in the case of transition magnetic moment $\Delta N$ we obtain a 
fairly good agreement. 
In order to compare the
present results with other recent similar calculations 
\cite{{song},{johan}}, in the table we have included  these results 
also.
A closer examination of the decuplet magnetic moments reveals several
interesting points which would have bearing on the 
generalized Cheng-Li mechanism. 
For example, in the case of $\Delta^-$ and $\Sigma^-$,
because the orbital part dominates over the ``quark sea''
polarization,  the magnetic moments are higher as compared to the 
results of NRQM and Refs. \cite{{song},{johan}}.
On the other hand, in the case of $\Delta^+$ and $\Sigma^+$, the
``quark sea'' polarization dominates over the orbital part as a consequence of
which the magnetic moment contribution is more or less the same as
that of the results of NRQM as well as those of Refs. \cite{{song},{johan}}.
In general, one can find that whenever there is an excess of $d$ quarks
the orbital part dominates, whereas when we have an excess of $u$ quarks, 
the ``quark sea'' polarization dominates.
A measurement of these magnetic moments, therefore, would have
important implications for the $\chi$QM as well as the 
Cheng-Li mechanism with its generalization.

While carrying out the fit, as mentioned earlier, the quark masses
which have been employed for the calculations correspond to the
generally accepted values used for hadron spectroscopic calculations. 
It may be of interest to study the
variation of these masses on the magnetic moments. To this end, in
Table \ref{masses}, we have investigated the effect of varying
valence quark masses. As is evident from the table we find that
results worsen in both the cases, for example, when they are reduced or
increased compared to the ones considered earlier.  The violation
of CGSR is also fitted best for the generally accepted mass values 
employed in our
calculations. These results remain true for E866 as well as the NMC data.
This looks to be surprising as the hadron spectroscopic predictions are
known to be somewhat insensitive to the  valence quark masses.

While discussing the inputs, we have already seen that
$\chi$QM$_{gcm}$ is able to give an excellent fit to the $Q^2$ independent
flavor non-singlet components, for example, $\Delta_3$ and $\Delta_8$. 
The flavor singlet component $\Delta \Sigma$ is also known to be
having a weak $Q^2$ dependence \cite{chengspin,anomaly}, 
therefore in principle we should be
able to get a good fit to this quantity also. However, in 
the absence of gluon contribution, expectedly
the agreement does not turn out to
be as impressive as in the case of flavor non-singlet components. 
The quark spin distribution functions can  be corrected by 
the inclusion of gluon anomaly \cite{chengspin,anomaly} through
\be
\Delta q(Q^2)= \Delta q- 
 \frac{\alpha_s(Q^2)}{2 \pi} \Delta g(Q^2)\,,
\ee
therefore, the flavor singlet component
of the total helicity in the $\chi$QM can be expressed as 
\be
\Delta \Sigma(Q^2)= \Delta \Sigma- 
 \frac{3\alpha_s(Q^2)}{2 \pi} \Delta g(Q^2)\,, \label{gluona}
\ee
where $\Delta \Sigma(Q^2)$ and $\Delta q(Q^2)$ are the experimentally measured 
quantities whereas $\Delta \Sigma$  and $\Delta q$ correspond to the 
calculated quantities in the $\chi$QM. Using 
$\Delta \Sigma(Q^2)=0.30\pm0.06$ \cite{abe},  $\Delta \Sigma=0.62$
and $\alpha_s(Q^2=5{\rm GeV}^2)= 0.287\pm0.020$ \cite{PDG}, the 
gluon polarization, $\Delta g(Q^2)$, comes out to be 2.33.
Interestingly, this value comes out to be in fair 
agreement with certain recent measurements \cite{hermes} 
as well as theoretical estimates \cite{{deltag},{bag}}. 
The effects of the gluon polarization can easily be incorporated into the
calculations of spin polarization functions and magnetic moments,
without getting into the details, the calculated values of the 
relevant phenomenological quantities affected by the gluon polarizations
are presented in Table \ref{gluon}.
From the table, we find that the present value of $\Delta g$
improves the results of various quantities, in particular, 
for $\Delta u$,  $\Delta d$, $\Delta s$, $\Delta \Sigma$, 
$\mu_n$, $\mu_{\Sigma^-}$,  $\mu_{\Lambda}$ and
$\mu_{\Sigma \Lambda}$ the
results hardly leave anything to be desired whereas $\mu_{\Xi^-}$, a
difficult case in most of the models, also registers a good deal of 
improvement. 
The decuplet magnetic moments do not show much change when correction
due to $\Delta g$ are included, for example, in the case of $\Omega^-$, 
$-2.01$ changes to $-2.04$ whereas in the case of $\Delta^{++}$,  $5.97$ 
changes to $5.94$. In the absence of experimental data for the other 
decuplet baryons, we have not included the $\Delta g$ corrected results 
in the table.

It may be of interest to emphasize here that the excellent fit achieved 
for the spin distribution functions, quark distribution functions and 
hyperon parameters alongwith the magnetic moments as well as the gluon 
polarization, strongly suggests a deeper significance of the values of 
the parameters employed, in particular the quark masses and the mixing 
angle.

\section{Summary and conclusion} \label{summary}
To summarize, the input parameters 
pertaining to the $\chi$QM with and without configuration mixing,
have been fixed by carrying out a brief analysis incorporating the 
latest data pertaining to $\bar u-\bar d$ asymmetry and 
spin polarization functions.
These parameters of $\chi$QM when used with the generally accepted values 
of the quark masses $M_q$,
incorporating the ``quark sea'' contribution as well as its orbital 
angular momentum through the generalized Cheng-Li mechanism, not only 
improve the baryon magnetic moments as compared to NRQM but also give
a non zero value for $\Delta$CG. 
The predictions of the  $\chi$QM with the generalized 
Cheng-Li mechanism improve further
when effects of configuration mixing and ``mass adjustments'' due to
confinement effects are included, for example, in the case of E866 data we get
an excellent fit for the octet magnetic moments and an almost perfect
fit for $\Delta$CG. 
Interestingly, we find that generalized 
Cheng-Li mechanism coupled with the effects
of configuration mixing plays a crucial role in fitting the individual 
magnetic moments, whereas ``mass adjustments'' alongwith the 
generalized Cheng-Li mechanism
play an important role in fitting $\Delta$CG.
When the above analysis is repeated with the earlier NMC data, a similar
level of agreement is obtained however the results in the
case of E866 look to be better.
Interestingly, we find that the masses $M_u=M_d=330$ MeV, after
corrections due to configuration mixing and ``mass adjustments'', 
provide the best fit for the magnetic moments.

In the case of decuplet baryon magnetic moments, we find a good
agreement of $\Delta^{++}$ and $\Omega^-$ with the experimental data. 
On comparison of our results with the corresponding results of 
Song and Linde {\it et al.}, we
find that the measurement of the $\Delta^+$, $\Delta^-$, $\Sigma^+$,
$\Sigma^-$ would have implications for the Cheng-Li mechanism.

Within $\chi$QM with configuration mixing, when $\Delta q(Q^2)$ is 
corrected by the inclusion of  gluon contribution through axial 
anomaly \cite{{chengspin},{anomaly}},
we not only obtain improvement in the quark spin 
distribution functions and magnetic moments but also the gluon 
polarization found in this manner is very much in  
agreement with certain recent measurements \cite{hermes} as 
well as  theoretical estimates \cite{{deltag},{bag}}.

In conclusion, we would like to state that the success of $\chi$QM 
with the Cheng-Li mechanism and configuration mixing in achieving an
excellent agreement regarding spin distribution functions, 
quark distribution functions and magnetic moments, 
strongly suggests that, at leading order, constituent quarks and the 
weakly interacting Goldstone bosons constitute the appropriate degrees 
of freedom in the nonperturbative regime of QCD with the weakly 
interacting gluons ({\it a la} Manohar and Georgi) providing 
the first order corrections.

\vskip .2cm
 {\bf ACKNOWLEDGMENTS}\\
The authors would like to thank S.D. Sharma, M. Randhawa and J.M.S. Rana
for a few useful discussions.
H.D. would like to thank CSIR, Govt. of India, for
 financial support and the chairman,
 Department of Physics, for providing facilities to work
 in the department.

\appendix
\renewcommand{\theequation}{A-\arabic{equation}}
  \setcounter{equation}{0} 

\begin{center}
{\bf APPENDIX A} 
\end{center}

The magnetic moment of a given baryon in $\chi$QM with sea and orbital
contributions, following Eq. (\ref{totalmag}), is given as 
\be
\mu(B)_{{\rm total}} = \mu(B)_{{\rm val}} + \mu(B)_{{\rm sea}} +
\mu(B)_{{\rm orbit}}\,. \label{totalmagmom}
\ee
To calculate the valence contribution to the magnetic moment,
$\mu(B)_{{\rm val}}$, we first express it in terms of
valence quark polarizations ($\Delta q_{{\rm val}}$) 
and the quark magnetic moments ($\mu_q$), for example,
\be
 \mu(B)_{{\rm val}}=\Delta u_{{\rm val}}\mu_u +\Delta d_{{\rm val}}\mu_d +
\Delta s_{{\rm val}}\mu_s. \label{pval}
\ee
The quark polarizations can be calculated from the spin
structure of a given baryon. Using Eqs. (\ref{BNB}) 
and (\ref{mixed}) of the text, the spin structure of a baryon in the 
``mixed'' octet is given as 
\be
\hat B \equiv \langle B|{\cal N}|B \rangle={\cos}^2 \phi
{\langle 56,0^+|{\cal N}|56,0^+\rangle}_B +
{\sin}^2 \phi {\langle 70,0^+|{\cal N}|70,0^+ \rangle}_B\,. \label{spin st}
\ee
For the case of proton, using Eqs. (\ref{56}) and
(\ref{70}) of the text, we have 
\bea 
{\langle 56,0^+|{\cal N}|56,0^+ \rangle}_p&=&\frac{5}{3} u_{+} +\frac{1}{3} u_{-}+
\frac{1}{3} d_{+} +\frac{2}{3} d_{-}\,, \label{56proton} \\ 
 {\langle 70,0^+|{\cal N}|70,0^+ \rangle}_p&=&\frac{4}{3} u_{+} +\frac{2}{3} u_{-}+
\frac{2}{3} d_{+} +\frac{1}{3} d_{-}\,. \label{70proton}
\eea
The valence contribution to the magnetic moment for the proton,
$\mu(p)_{{\rm val}}$, can be found by using 
Eqs.  (\ref{pval}), (\ref{spin st}), (\ref{56proton}) and 
(\ref{70proton}), for example,
\be
 \mu(p)_{{\rm val}}=\left[{\cos}^2 \phi \left(\frac{4}{3} \right)
+{\sin}^2 \phi \left(\frac{2}{3} \right) \right ]\mu_u +
\left [{\cos}^2 \phi \left(-\frac{1}{3} \right)  +
  {\sin}^2 \phi \left(\frac{1}{3}  \right) \right]\mu_d +
[0]\mu_s\,. \label{val p} 
\ee
For the $\Lambda$ hyperon, we have
\bea
{ \langle 56,0^+|{\cal N}|56,0^+ \rangle}_{\Lambda}&=&\frac{1}{2} u_{+} +\frac{1}{2} u_{-}+
\frac{1}{2} d_{+} +\frac{1}{2} d_{-}+
1 s_{+} +0 s_{-}\,, \label{56lambda} \\
{ \langle 70,0^+|{\cal N}|70,0^+ \rangle}_{\Lambda}&=&\frac{2}{3} u_{+} +\frac{1}{3} u_{-}+
\frac{2}{3} d_{+} +\frac{1}{3} d_{-}+
\frac{2}{3} s_{+} +\frac{1}{3} s_{-}\,, \label{70lambda}
\eea
and  
\be
\mu(\Lambda)_{{\rm val}}=\left[{\sin}^2 \phi \left(\frac{1}{3}
\right) \right]\mu_u 
+ \left[{\sin}^2 \phi \left(\frac{1}{3}  \right)\right]\mu_d +
\left[{\cos}^2 \phi (1) +{\sin}^2 \phi \left(\frac{1}{3} \right)
\right]\mu_s\,.  \label{val lambda} 
\ee
Similarly, we can calculate the valence contribution to the magnetic 
moments for other octet baryons, however the calculation
of the transition magnetic moment $\mu(\Sigma \Lambda)$ is somewhat
different, for which we have
\bea 
{\langle 56,0^+|{\cal N}|56,0^+ \rangle}_{\Sigma \Lambda}&=&\frac{1}{2 \sqrt 3} u_{+} -\frac{1}{\sqrt 3} u_{-}-
\frac{1}{2 \sqrt 3} d_{+} +\frac{1}{\sqrt 3} d_{-}\,, \label{56tran} \\ 
 {\langle 70,0^+|{\cal N}|70,0^+ \rangle}_{\Sigma \Lambda}&=&
\frac{1}{4 \sqrt 3} u_{+} +\frac{3}{4 \sqrt 3} u_{-}-
\frac{1}{4 \sqrt 3} d_{+} -\frac{3}{4 \sqrt 3} d_{-}\,, \label{70tran}
\eea
giving
\be
\mu(\Sigma \Lambda)_{{\rm val}}=-\frac{1}{2 \sqrt 3}\left[\left (\cos^2 \phi(-\frac{1}{\sqrt 3})+\sin^2 \phi(-\frac{1}{2 \sqrt 3}) \right)-
2 \left( \cos^2 \phi(\frac{1}{\sqrt 3})+\sin^2 \phi(\frac{1}{2 \sqrt 3})\right)\right](\mu_u-\mu_d)\,. \label{val tran}
\ee
The ``quark sea'' contribution to the magnetic moment of 
a given baryon, $\mu(B)_{{\rm sea}}$, can be expressed in terms of the 
sea quark polarizations ($\Delta q_{{\rm sea}}$) and $\mu_q$ as
\be
 \mu(B)_{{\rm sea}}=\Delta u_{{\rm sea}}\mu_u+\Delta d_{{\rm sea}}\mu_d+
\Delta s_{{\rm sea}}\mu_s\,. \label{seacont}
\ee
To calculate $\Delta q_{{\rm sea}}$ for different quarks in a given baryon, 
we consider the spin structure of the baryon alongwith the ``quark sea''.
Using Eq. (\ref{q}) of the text and Eqs. (\ref{spin st}), (\ref{56proton}) and 
(\ref{70proton}), the spin structure of the proton and the associated 
``quark sea'' is given as

\[ \hat p={\cos}^2 \phi \left [ \frac{5}{3}(\sum P_u u_{+} + 
|\psi(u_{+})|^2)+
\frac{1}{3}(\sum P_u u_{-} + |\psi(u_{-})|^2)+
\frac{1}{3}(\sum P_d d_{+} + |\psi(d_{+})|^2) + \frac{2}{3}(\sum P_d d_{-} + 
|\psi(d_{-})|^2) \right ] \]
\be
+{\sin}^2 \phi \left [ \frac{4}{3}(\sum P_u u_{+} + 
|\psi(u_{+})|^2)+
\frac{2}{3}(\sum P_u u_{-} + |\psi(u_{-})|^2)+ 
\frac{2}{3}(\sum P_d d_{+} + |\psi(d_{+})|^2)+
\frac{1}{3}(\sum P_d d_{-} + |\psi(d_{-})|^2) \right ], \label{hat p}
\ee

where
\[ \sum P_u= a\left( \frac{9+\beta^2+2 \zeta^2}{6} +\alpha^2\right)~~~
{\rm and}~~~
|\psi(u_{\pm})|^2=\frac{a}{6}(3+\beta^2+2 \zeta^2)u_{\mp}+
a d_{\mp}+a \alpha^2 s_{\mp}\,,      \]

\[ \sum P_d= a\left( \frac{9+\beta^2+2 \zeta^2}{6} +\alpha^2\right)~~~
{\rm and}~~~ 
|\psi(d_{\pm})|^2=a u_{\mp}+
\frac{a}{6}(3+\beta^2+2 \zeta^2)d_{\mp}+ a \alpha^2 s_{\mp}\,, 
                                               \]

\[ \sum P_s= a\left( \frac{2 \beta^2+\zeta^2}{3}+2 \alpha^2\right)~~~
{\rm and}~~~
|\psi(s_{\pm})|^2=   a \alpha^2 u_{\mp}+
a \alpha^2 d_{\mp}+\frac{a}{3}(2 \beta^2+\zeta^2)s_{\mp}\,.
 \]
Using Eqs. (\ref{seacont}) and (\ref{hat p}), the ``quark sea'' 
contribution to the magnetic moment for the case of proton is given as
\bea
 \mu(p)_{{\rm sea}}&=&\left[-{{\cos}}^2 \phi \left(\frac{a}{3} 
(7+4 \alpha^2+
 \frac{4}{3}\beta^2 +\frac{8}{3} \zeta^2)\right)-{\sin}^2 \phi
\left(\frac{a}{3} (5+2 \alpha^2 
+\frac{2}{3}\beta^2 +\frac{4}{3} \zeta^2)\right)\right]\mu_u \nonumber \\
+& &\left[-{\cos}^2 \phi \left(\frac{a}{3} (2-\alpha^2
-\frac{1}{3}\beta^2 -\frac{2}{3} \zeta^2)\right)-{\sin}^2 \phi \left(\frac{a}{3} (4+\alpha^2
+\frac{1}{3}\beta^2 +\frac{2}{3} \zeta^2)\right)\right]\mu_d+
\left[-a \alpha^2\right]\mu_s\,. \label{sea p}
\eea
Similarly, the spin structure for $\Lambda$ can obtained by substituting 
 Eq. (\ref{q})  in Eqs. (\ref{56lambda}) and 
(\ref{70lambda}) and  is given as

\bea
 \hat \Lambda&=&{\cos}^2 \phi \left [ \frac{1}{2}\left ( \sum P_u u_{+} + 
|\psi(u_{+})|^2 +
\sum P_u u_{-} + |\psi(u_{-})|^2+
\sum P_d d_{+} + |\psi(d_{+})|^2+\sum P_d d_{-} + |\psi(d_{-})|^2 \right )
 \right .  \nonumber \\
 &+& \left .\sum P_s s_{+} +|\psi(s_{+})|^2 \right ] 
 + {\sin}^2 \phi \left [ \frac{2}{3} \left (\sum P_u u_{+}
+ |\psi(u_{+})|^2+\sum P_d d_{+} + |\psi(d_{+})|^2+
\sum P_s s_{+} +|\psi(s_{+})|^2 \right )  \right .\nonumber \\
&+& \left . 
\frac{1}{3} \left (\sum P_u u_{-} + 
|\psi(u_{-})|^2 + \sum P_d d_{-} + |\psi(d_{-})|^2+
\sum P_s s_{-} +|\psi(s_{-})|^2 \right )  \right ]. \label{hat lambda}
\eea
The  ``quark sea'' contribution to the magnetic moment for 
the case of $\Lambda$ is given  as
\bea
 \mu(\Lambda)_{{\rm sea}}&=&
\left[-{\cos}^2 \phi \left(a \alpha^2\right)-{\sin}^2 \phi 
\left(\frac{a}{9}(9+6 \alpha^2+\beta^2+ 2
\zeta^2)\right)\right]\mu_u +
\left[-{\cos}^2 \phi \left(a \alpha^2\right)-{\sin}^2 \phi 
\left(\frac{a}{9}(9+6 \alpha^2+\beta^2+ 2
\zeta^2)\right)\right]\mu_d \nonumber \\
&+&\left[-{\cos}^2 \phi \left(\frac{a}{3}(6 \alpha^2+
4 \beta^2 +2 \zeta^2)\right)-{\sin}^2 \phi \left(\frac{4}{9}a(3
\alpha^2+2 \beta^2+ \zeta^2)\right)\right]\mu_s\,. \label{sea lambda}
\eea
Similarly, one can calculate the contribution of the 
``quark sea'' spin polarizations to the magnetic moments of the other baryons
and these have been listed in Table \ref{ocsea}.
For the transition magnetic moment $\mu(\Sigma \Lambda)$, the spin structure 
can be obtained from Eqs. (\ref{q}), (\ref{56tran}) and (\ref{70tran}) 
and is given as 

{\footnotesize\[ \widehat {\Sigma \Lambda}={\cos}^2 \phi \left [ 
\frac{1}{2 \sqrt 3}(\sum P_u u_{+} + |\psi(u_{+})|^2)-
\frac{1}{\sqrt 3}(\sum P_u u_{-} + |\psi(u_{-})|^2)-
\frac{1}{2\sqrt 3}(\sum P_d d_{+} + |\psi(d_{+})|^2) + 
\frac{1}{\sqrt 3}(\sum P_d d_{-} + |\psi(d_{-})|^2) \right ] \]
\be
+{\sin}^2 \phi \left [ \frac{1}{4 \sqrt3}(\sum P_u u_{+} + 
|\psi(u_{+})|^2)+
\frac{3}{4 \sqrt 3}(\sum P_u u_{-} + |\psi(u_{-})|^2)- 
\frac{1}{4 \sqrt 3}(\sum P_d d_{+} + |\psi(d_{+})|^2)-
\frac{3}{4 \sqrt 3}(\sum P_d d_{-} + |\psi(d_{-})|^2) \right ],
\ee}
giving the ``quark sea'' contribution to the transition magnetic moment as
\bea
 \mu(\Sigma \Lambda)_{{\rm sea}}&=&-\frac{1}{2 \sqrt 3}\left[
-{{\cos}}^2 \phi \left (\frac{a}{2 \sqrt 3} 
(3+3 \alpha^2+ \beta^2 +2 \zeta^2)\right )+{\sin}^2 \phi
\left (\frac{a}{2 \sqrt 3} (1+\alpha^2 
+\frac{1}{3}\beta^2 +\frac{2}{3} \zeta^2)\right)\right. \nonumber \\
&- &\left. 2 \left ( {\cos}^2 \phi \left (\frac{a}{2 \sqrt 3} (3+3 \alpha^2 +
\beta^2 + 2 \zeta^2)\right )-{\sin}^2 \phi \left (\frac{a}{2 \sqrt 3} (1+\alpha^2
+\frac{1}{3}\beta^2 +\frac{2}{3} \zeta^2)\right )\right)\right](\mu_u-\mu_d)\,.
 \label{sea tran}
\eea

For calculating the orbital contribution to the total magnetic moment, 
one has to use generalized 
Cheng-Li mechanism expressed in Eq. (\ref{orbit}) and for 
the case of proton and $\Lambda$ it is given as
\bea
 \mu(p)_{{\rm orbit}} &=&{\cos}^2 \phi\left [ \frac{4}{3} 
[\mu (u_+ \rightarrow)] - \frac{1}{3} [\mu (d_+ \rightarrow)] \right]+
{\sin}^2 \phi\left[ \frac{2}{3} [\mu (u_+ \rightarrow)]
+\frac{1}{3} [\mu (d_+ \rightarrow)] \right ],  \label{orbit p} \\
 \mu(\Lambda)_{{\rm orbit}} &=&{\cos}^2 \phi [ \mu (s_+
\rightarrow)]
+{\sin}^2 \phi\left[\frac{1}{3} [\mu (u_+ \rightarrow)]
+\frac{1}{3}[\mu (d_+ \rightarrow)+
\frac{1}{3}[\mu (s_+ \rightarrow) ] \right ]. \label{orbit lambda}
\eea
For the case of $\Sigma \Lambda$, the orbital contribution to 
the magnetic moment is
\be
 \mu(\Sigma \Lambda)_{{\rm orbit}} =\left[{\cos}^2 \phi(\frac{1}{2})
+{\sin}^2 \phi(\frac{1}{4}) \right ]
([\mu (u_+ \rightarrow)]-[\mu (d_+ \rightarrow)])\,. \label{orbit tran}
\ee

Using Eq. (\ref{totalmagmom}) one can calculate the total magnetic moment
of $p$, $\Lambda$ and $\Sigma \Lambda$. The magnetic moments
of other octet baryons can similarly be calculated.

As an example of the decuplet baryon, we detail below the calculation
of magnetic moment of $\Delta^{+}$.
In the absence of any mixing, the spin structure for $\Delta^{+}$,
using Eqs. (\ref{56decuplet}) of the text, is given as
\be
 {\langle56,0^+|{\cal N}|56,0^+\rangle}_{\Delta^{+}}=2 u_{+}+ d_{+}\,. 
\label{56delta}
\ee
The valence contribution to the total magnetic moment is expressed as  
\bea
\mu(\Delta^+)_{{\rm val}}&=&\Delta u_{{\rm val}}\mu_u +
\Delta d_{{\rm val}}\mu_d +\Delta
s_{{\rm val}}\mu_s \nonumber \\
& =& 2 \mu_u +1\mu_d+ 0 \mu_s\,.  \label{dec val}
\eea

The contribution of the ``quark sea''  to the total magnetic moment 
in terms of the ``quark sea'' polarizations and $\mu_q$ is expressed as
\be
 \mu(\Delta^+)_{{\rm sea}}=\Delta u_{{\rm sea}}\mu_u+
\Delta d_{{\rm sea}}\mu_d+\Delta s_{{\rm sea}}\mu_s\,. 
\ee
By substituting Eq. (\ref{q}) in Eq. (\ref{56delta}), we obtain the 
spin structure of $\Delta^+$ and the associated ``quark sea'' 
which is expressed as
\be
  \hat {\Delta^{+}}=2 (\sum P_u u_{+} + 
|\psi(u_{+})|^2)+
(\sum P_d d_{+} + |\psi(d_{+})|^2)\,,
\ee
and the ``quark sea'' contribution  to the magnetic moment is consequently 
given as
\be
 \mu(\Delta^+)_{{\rm sea}}=\left[- a( 5+2 \alpha ^2 + \frac{2}{3} \beta^2+\frac{4}{3} \zeta^2) \right]\mu_u+
\left[- a( 4 + \alpha ^2 + \frac{1}{3} \beta^2+\frac{2}{3}\zeta^2)
\right]\mu_d+\left[-3a \alpha^2 \right]\mu_s\,.   \label{dec sea}
\ee
The contribution of the ``quark sea'' to the magnetic moment of other
decuplet baryons can similarly be calculated in terms of the
``quark sea'' polarizations, the expressions for which are
given Table \ref{ocsea}.

The orbital contribution to the total magnetic moment, 
as given by Eq. (\ref{de orbit}),  is expressed as
\be
 \mu(\Delta^+)_{orbit} =2 [\mu (u_+ \rightarrow)] +
[\mu (d_+ \rightarrow)]\,. \label{3dec orbit}
\ee
Substituting Eqs. (\ref{dec val}), (\ref{dec sea}) and 
(\ref{3dec orbit}) in Eq. (\ref{totalmagmom}) we get the total 
magnetic moment of $\Delta^+$. We can also calculate the 
transition magnetic moment $\mu(\Delta N)$ in a similar manner as we have 
calculated $\mu(\Sigma \Lambda)$.

\pagebreak

\begin{table}
\begin{center}
\begin{tabular}{cccc}       
Baryons & $\Delta u_{{\rm sea}}$ &$\Delta d_{{\rm sea}}$ &  
$\Delta s_{{\rm sea}}$ \\ \hline 
$p(uud)$ &$-{\cos}^2 \phi \left[\frac{a}{3} (7+4 \alpha^2+
 \frac{4}{3}\beta^2 +\frac{8}{3} \zeta^2)\right]$
  &$-{\cos}^2 \phi \left[\frac{a}{3} (2-\alpha^2
-\frac{1}{3}\beta^2 -\frac{2}{3} \zeta^2)\right]$
 & $-a \alpha^2$ \\ 
 & $-{\sin}^2 \phi \left[\frac{a}{3} (5+2 \alpha^2
+\frac{2}{3}\beta^2 +\frac{4}{3} \zeta^2)\right]$
&$-{\sin}^2 \phi \left[\frac{a}{3} (4+\alpha^2
+\frac{1}{3}\beta^2 +\frac{2}{3} \zeta^2)\right]$
& \\

 $\Sigma^+(uus)$ & $-{\cos}^2 \phi \left[\frac{a}{3}
   (8+3 \alpha^2+ \frac{4}{3}\beta^2+ \frac{8}{3} \zeta^2)\right]$ & 
$-{\cos}^2 \phi \left[\frac{a}{3}(4-\alpha^2) \right]$
 & $-{\cos}^2 \phi \left[\frac{a}{3} (2
\alpha^2-\frac{4}{3}\beta^2 -\frac{2}{3} \zeta^2)\right]$ \\
 &$-{\sin}^2 \phi \left[\frac{a}{3} (4+3 \alpha^2+
 \frac{2}{3}\beta^2 +\frac{4}{3} \zeta^2)\right]$ 
& $-{\sin}^2 \phi \left[\frac{a}{3}(2+\alpha^2) \right]$
& $-{\sin}^2 \phi  \left[\frac{a}{3} (4 \alpha^2+
 \frac{4}{3}\beta^2 +\frac{2}{3} \zeta^2)\right]$ \\ 

 $\Xi^o(uss)$ &$-{\cos}^2 \phi \left[\frac{a}{3}
   (3 \alpha^2-2- \frac{1}{3}\beta^2 -\frac{2}{3} \zeta^2)\right]$
&$-{\cos}^2 \phi \left[\frac{a}{3}(4\alpha^2-1) \right]$
&$-{\cos}^2 \phi \left[\frac{a}{3} (7 \alpha^2+
\frac{16}{3}\beta^2 +\frac{8}{3} \zeta^2)\right]$ \\
&  $-{\sin}^2 \phi \left[\frac{a}{3} (2+3 \alpha^2+
\frac{1}{3}\beta^2 +\frac{2}{3} \zeta^2)\right]$
&$ -{\sin}^2 \phi \left[\frac{a}{3}(1+2 \alpha^2) \right] $
& $-{\sin}^2 \phi \left[\frac{a}{3} (5 \alpha^2+
\frac{8}{3}\beta^2 +\frac{4}{3} \zeta^2)\right]$\\

$\Lambda(uds)$&$-{\cos}^2 \phi \left[a \alpha^2\right]$ &
$-{\cos}^2 \phi \left[a \alpha^2\right]$ &
$-{\cos}^2 \phi \left[\frac{a}{3}(6 \alpha^2+
4 \beta^2 +2 \zeta^2)\right]$ \\
&$-{\sin}^2 \phi \left[\frac{a}{9}(9+6 \alpha^2+\beta^2+
2 \zeta^2)\right]$ &
$-{\sin}^2 \phi \left[\frac{a}{9}(9+6 \alpha^2+\beta^2+
2 \zeta^2)\right]$ &
$-{\sin}^2 \phi \left[\frac{4}{9}a(3 \alpha^2+2 \beta^2+
\zeta^2)\right]$ \\ 

$\Sigma \Lambda$ & $-{\cos}^2 \phi \left[\frac{a}{2 \sqrt 3}(3+3 \alpha^2+\beta^2+2 \zeta^2)\right]$ &$ {\cos}^2 \phi \left[\frac{a}{2 \sqrt 3}(3+3 \alpha^2+\beta^2+2 \zeta^2)\right]$ &0 \\
&${\sin}^2 \phi \left[\frac{a}{2 \sqrt 3}(1+\alpha^2+\frac{\beta^2}{3}+\frac{2}{3} \zeta^2)\right]$ &$-{\sin}^2 \phi \left[\frac{a}{2 \sqrt 3}(1+\alpha^2+\frac{\beta^2}{3}+\frac{2}{3} \zeta^2)\right]$ & \\
\hline

$\Delta^{++}(uuu)$ & $- a( 6 + 3 \alpha ^2 +\beta^2+2 \zeta^2)$  & $-3a$ &  $-3a \alpha^2$ \\  
$\Delta^{+}(uud)$ & $- a( 5+2 \alpha ^2 + \frac{2}{3} \beta^2+\frac{4}{3} \zeta^2)$  &
   $- a( 4 + \alpha ^2 + \frac{1}{3} \beta^2+\frac{2}{3}\zeta^2)$ & $-3a \alpha^2$ \\  

$\Sigma^{*^+}(uus)$ & $- a( 4+3 \alpha^2 + \frac{2}{3} \beta^2+\frac{4}{3}\zeta^2)$  &
 $- a( \alpha ^2 + 2)$ & $- 2 a( 2 \alpha ^2 + \frac{2}{3} \beta^2+\frac{1}{3} \zeta^2)$  \\
$\Sigma^{*^o}(uds)$ & $- a( 3 + 2\alpha ^2 +\frac{1}{3} \beta^2+\frac{2}{3} \zeta^2)$ & 
$- a( 3 + 2\alpha ^2 +\frac{1}{3} \beta^2+\frac{2}{3} \zeta^2)$ & 
 $- 2a( 2\alpha ^2 +\frac{2}{3} \beta^2+\frac{1}{3} \zeta^2)$ \\

$\Xi^{*^o}(uss)$  &  $-a (2+3 \alpha^2+\frac{1}{3}\beta^2+\frac{2}{3}\zeta^2)$  &  
$-a(2 \alpha^2 +1)$ & $- a(5 \alpha^2 + \frac{8}{3} \beta^2+\frac{4}{3} \zeta^2)$ \\

$\Omega^{-}(sss)$ &  $-3a \alpha^2$ &  $-3a \alpha^2$ & 
$- 6a(\alpha^2 + \frac{2}{3} \beta^2+\frac{1}{3} \zeta^2)$ \\   
$\Delta N$ & $-\frac{2}{3\sqrt 6}(3+3 \alpha^2+\beta^2+2\zeta^2)$&
$\frac{2}{3\sqrt 6}(3+3 \alpha^2+\beta^2+2\zeta^2)$& 0 \\

\end{tabular}
\end{center}
\caption{Sea quark spin polarizations for the ``mixed'' 
octet baryons and decuplet baryons 
in terms of the $\chi$QM parameters $a$, $\alpha$, $\beta$ and $\zeta$ as 
discussed in the text. The spin polarizations
for the other baryons can be found from isospin  symmetry.
The spin structure of the octet baryon $B$ without
configuration mixing can be obtained by taking $\phi=0$.} 
\label{ocsea}
\end{table}

\begin{table}
\begin{center}
\begin{tabular}{cccccc}     
              
 &  & \multicolumn{2}{c}{$\chi$QM} & 
\multicolumn{2}{c} {$\chi$QM$_{gcm}$} \\  
Parameter & Data & \multicolumn{2}{c}{$\alpha=0.6,\beta=0.9 $} & 
\multicolumn{2}{c} {$\alpha=0.4 ,\beta=0.7$} \\ \cline{3-6} 
&&NMC &E866 & NMC & E866 \\ \hline

 $\Delta$ u & 0.85 $\pm$ 0.05 {\cite{adams}} & 0.88 & 0.92  & 0.91 & 0.95 \\

 $\Delta$ d & -0.41  $\pm$ 0.05  {\cite{adams}} & -0.35  &-0.36 & -0.33 & -0.31 \\

$\Delta$ s &-0.07  $\pm$ 0.05  {\cite{adams}} & -0.05 & -0.05 & -0.02 & -0.02 \\
$\Delta_3$ & 1.267  $\pm$ .0035  {\cite{PDG}} & 1.23 & 1.28  & 1.24 & 1.26 \\
$\Delta_8$ & 0.58$\pm$0.025 {\cite{adams}} & 0.63 & 0.66  & 0.61 & 0.67 \\
$\bar u-\bar d$ & -0.147 $\pm$ .024 {\cite{NMC}}  &
0.147 & 0.12 & 0.147 & 0.12  \\
                & -0.118 $\pm$ .015 \cite{E866} &&&& \\

$\bar d/\bar u$ & 1.96 $\pm$ 0.246 {\cite{baldit}} &  1.89 & 1.59  &
1.89 &  1.59 \\
                & 1.41 $\pm$ 0.146 \cite{E866} && && \\

$f_s$ &  0.10 $\pm$ 0.06 {\cite{ao}} & 0.15 & 0.13 & 0.07 & 0.05  \\

$f_3/f_8$ & 0.21 $\pm$ 0.05 {\cite{cheng}} & 0.25 & 0.25 &
0.21 &   0.21 \\     

\end{tabular}
\end{center}
\caption{$\chi$QM parameters (with and without configuration mixing) 
obtained after fitting spin and quark distribution functions.
$\chi$QM$_{gcm}$ corresponds to the case
where the ``mixed'' nucleon (Eq. (\ref{mixed}))
has been used with the mixing angle $\phi=20^o$.} \label {ud}
\end{table}

\begin{table}
\begin{center}
\begin{tabular}{ccccccc}      
Parameter &$\phi$ &$a$ &$\alpha$ &$\beta$ &$\zeta_{E866}$ & $\zeta_{NMC}$ \\  
\hline
Value & $20^o$ & 0.1 & 0.4 & 0.7 & $-0.3-\beta/2$ & $-0.7-\beta/2$ \\

\end{tabular}
\end{center}
\caption{ Input values of various parameters used in the analysis.} 
\label{input}
\end{table}

\begin{table}
{\footnotesize
\begin{center}
\begin{tabular}{ccccccccccccccc}   
 &  & & \multicolumn{4}{c} { } & \multicolumn{4}{c} {} & 
\multicolumn{4}{c} {$\chi$QM with mass }\\  

&  & & \multicolumn{4}{c} {$\chi$QM} & \multicolumn{4}{c}
{$\chi$QM with} &  \multicolumn{4}{c} {adjustments and}\\ 
&  & & \multicolumn{4}{c} {} & \multicolumn{4}{c}
{mass adjustments} &  \multicolumn{4}{c} {configuration
mixing}\\ 
\cline{4-7}  \cline{8-11} \cline {12-15}

Octet & Data & NRQM &  Valence  & Sea & Orbital &  Total &  
Valence  & Sea & Orbital &  Total &  Valence  & Sea & Orbital &  Total \\
 baryons & ~\cite{PDG} &  &   & &&&&&&&&&& \\  \hline

p          & 2.79$\pm$0.00  & 2.72 & 3.00 & -0.70 & 0.54 & 2.84 &  
3.17   & -0.59&    0.45 & 3.03 &  
2.94    & -0.55& 0.41 & 2.80   \\
n          & -1.91$\pm$0.00 & -1.81 & -2.00 & 0.34 & -0.41 & -2.07 & 
-2.11    & 0.24  & -0.37 & -2.24 & 
 -1.86  &  0.20 & -0.33  & -1.99   \\

$\Sigma^-$ & -1.16$\pm$0.025 &  -1.01 & -1.12 & 0.13 & -0.29 & -1.28 & 
-1.08 & 0.08& -0.26 &   -1.26 &
   -1.05    & 0.07 &  -0.22 & -1.20  \\
$\Sigma^+$ & 2.45$\pm$0.01  & 2.61 & 2.88 & -0.69 & 0.45 & 2.64 &  
2.80  &  -0.55 &  0.37 & 2.62 & 
 2.59 & -0.50 & 0.34 &  2.43  \\

$\Xi^o$    & -1.25$\pm$0.014 & -1.41 & -1.53 & 0.37 & -0.23 & -1.39 & 
-1.53 & 0.22 &  -0.16 &  -1.47 & 
 -1.32 & 0.21 &-0.13 & -1.24  \\
$\Xi^-$    & -0.65$\pm$0.002 & -0.50 & -0.53 & 0.09 & -0.06 & -0.50 & 
-0.59 & 0.06  & -0.01 & -0.54 &  
 -0.61 & 0.06&  -0.01 &   -0.56  \\

$\Lambda$ & -0.61$\pm$0.004  & -0.59 & -0.65 & 0.10 & -0.08 & -0.63 & 
-0.69  & 0.05  & -0.04  &-0.68  &
-0.59 & 0.04 &-0.04  & -0.59   \\
$\Sigma \Lambda$ & 1.61$\pm$0.08 & 1.51 & 1.41 & -0.02 & 0.30& 1.69&
1.45 & -0.03 & 0.30   & 1.72   &
1.37  & -0.04 & 0.26 & 1.63    \\
 \hline
$\Delta$CG  & 0.49 $\pm$ 0.05 & 0 &  & & & 0.10 & 
& &&  0.46 &  & &&     0.48 \\  

\end{tabular}
\end{center}}
\caption{ Octet baryon magnetic moments in units of $\mu_N$ for the latest E866 data.}  \label{E866}
\end{table}

\begin{table}
{\footnotesize
\begin{center}
\begin{tabular}{ccccccccccccccc}     
 &  & & \multicolumn{4}{c} { } & \multicolumn{4}{c} {} & 
\multicolumn{4}{c} {$\chi$QM with mass }\\  

&  & & \multicolumn{4}{c} {$\chi$QM} & \multicolumn{4}{c}
{$\chi$QM with} &  \multicolumn{4}{c} {adjustments and}\\ 
&  & & \multicolumn{4}{c} {} & \multicolumn{4}{c}
{configuration mixing} &  \multicolumn{4}{c} {configuration
mixing}\\ 
\cline{4-7}  \cline{8-11} \cline {12-15}

Octet & Data & NRQM &  Valence  & Sea & Orbital &  Total &  
Valence  & Sea & Orbital &  Total &  Valence  & Sea & Orbital &  Total \\
 baryons & ~\cite{PDG} &  &   & &&&&&&&&&& \\  \hline

p          & 2.79$\pm$0.00  & 2.72 & 3.00 & -0.79 & 0.53 & 2.74 &  
2.76   & -0.62&    0.48 & 2.62 &  
2.94    & -0.65& 0.41 & 2.70   \\
n          & -1.91$\pm$0.00 & -1.81 & -2.00 & 0.30 & -0.29 & -1.99 & 
-1.76    & 0.25  & -0.39 & -1.90 & 
 -1.86  &  0.27 & -0.34  & -1.93   \\

$\Sigma^-$ & -1.16$\pm$0.025 &  -1.01 & -1.12 & 0.16 & -0.30 & -1.26 & 
-1.09 & 0.10& -0.25 &   -1.24 &
   -1.05    & 0.14 &  -0.26 & -1.17  \\
$\Sigma^+$ & 2.45$\pm$0.01  & 2.61 & 2.88 & -0.77 & 0.43 & 2.54 &  
2.67  &  -0.65 &  0.40 & 2.42 & 
 2.59 & -0.59 & 0.36 &  2.36  \\

$\Xi^o$    & -1.25$\pm$0.014 & -1.41 & -1.53 & 0.45 & -0.21 & -1.29 & 
-1.32 & 0.26 &  -0.16 &  -1.22 & 
 -1.32 & 0.26 &-0.14 & -1.20  \\
$\Xi^-$    & -0.65$\pm$0.002 & -0.50 & -0.53 & 0.08 & -0.01 & -0.46 & 
-0.56 & 0.09  & -0.01 & -0.48 &  
 -0.61 & 0.09&  -0.02 &   -0.54  \\

$\Lambda$ & -0.61$\pm$0.004  & -0.59 & -0.65 & 0.12 & -0.07 & -0.60 & 
-0.56  & 0.07  & -0.05  &-0.54  &
-0.59 & 0.07 &-0.05  & -0.57   \\
$\Sigma \Lambda$ & 1.61$\pm$0.08 & 1.51 & 1.41 & -0.01 & 0.31&1.71 &
1.41 & -0.01   & 0.26    & 1.66  &
1.37 & -0.02 & 0.26  & 1.61   \\
 \hline
$\Delta$CG  & 0.49 $\pm$ 0.05 & 0 &  & & & 0.10 & 
& &&  0.12 &  & &&     0.44 \\ 

\end{tabular}
\end{center}}
\caption{ Octet baryon magnetic moments in units of $\mu_N$ for the NMC data.}
 \label{NMC}
\end{table}

\begin{table}
{\footnotesize
\begin{center}
\begin{tabular}{cccccccccccc}       
              
Decuplet & Data & NRQM & X. Song  & Linde {\it et al.}  &  Valence  &
\multicolumn{2}{c}{Sea} &  
\multicolumn{2}{c}{Orbital} & \multicolumn{2}{c}{Total} \\ \cline{7-12}

 baryons & ~\cite{PDG} & &~{\cite{song}} & ~{\cite{johan}} &  & NMC &E866 &NMC
 &E866 &NMC & E866 \\ \hline 

$\Delta^{++}$ & 3.7 $<$ $\mu_{\Delta^{++}}$ $<$ 7.5& 5.43 & 5.55 &
5.21 & 6.36 & -1.59 & -1.31 & 0.94 & 0.92  & 5.71 & 5.97   \\
$\Delta^{+}$  & - & 2.72 & 2.73 & 2.45& 3.18 & -0.94 & -0.79 & 0.38 & 0.37 
& 2.62 & 2.76   \\
$\Delta^{o}$  & - & 0 & -0.09 & -0.30 & 0 & -0.28 & -0.28 & -0.18 & -0.18 
& -0.46 & -0.46  \\
$\Delta^{-}$  & - & -2.72 & -2.91 & -3.06 & -3.18 & 0.37 &  0.23 &
-0.74  & -0.73  & -3.55 & -3.68  \\

$\Sigma^{*+}$  & - & 3.02 & 3.09 & 2.85 & 3.24 &  -0.88 & -0.73 &  0.58 & 0.56 
& 2.94  & 3.07  \\ 
$\Sigma^{*o}$  & - & 0.30 & 0.27 & 0.09 & 0.33 & -0.28 &-0.26 & 0.01 & 0.01
& 0.06 & 0.08  \\
$\Sigma^{*-}$  & - & -2.41 & -2.55 & -2.66 & -2.58 & 0.32  & 0.20 &
-0.54 & -0.54  & -2.80 & -2.92  \\

$\Xi^{*o}$     & - & 0.60 & 0.63 & 0.49 & 0.52 & -0.27 & -0.24 & 0.21 & 0.21 
& 0.46  & 0.49  \\
$\Xi^{*-}$     & - & -2.11 & -2.19 & -2.27 & -2.30 & 0.31  & 0.21 &
-0.35 & -0.34  & -2.33  & -2.43  \\
$\Omega^{-}$  & -2.02 $\pm$ 0.005 & -1.81 & -1.83 & -1.87 & -2.07 &
0.30 & 0.21 & -0.14 &  -0.15 & -1.91 & -2.01  \\          
$\Delta N$ & 3.23$\pm$0.10$^*$ & 2.44 &-&-& 2.60 &-0.53& -0.41 & 0.46 &0.44 & 
2.53 & 2.63  \\

\end{tabular}
\end{center}}
{\footnotesize * pertains to the PDG 1994 data.}
\caption{Decuplet magnetic moments in units of $\mu_N$  for NMC and E866 data.} \label{decuplet}
\end{table}

\begin{table}
\begin{center}
\begin{tabular}{ccccccccc}      
              
Octet  & Data & NRQM &\multicolumn{2}{c} {$M_u,M_d=310$ MeV} & 
\multicolumn{2}{c} {$M_u,M_d=340$ MeV}& 
\multicolumn{2}{c} {$M_u,M_d=330$ MeV} \\ \cline{4-9}   
baryons& ~\cite{PDG} & & NMC &E866 &NMC & E866 & NMC & E866 \\ \hline
p          & 2.79$\pm$0.00  &  2.72 & 2.48 & 2.60 & 2.69 & 2.84 &   2.70 & 2.80   \\
n          & -1.91$\pm$0.00 &  -1.81 &  -1.79 & -1.88 & -1.96 & -2.06  &   
-1.93 &  -1.99    \\

$\Sigma^-$ & -1.16$\pm$0.025 &  -1.01  & -1.16& -1.20 & -1.28 & -1.32 &   
-1.17&  -1.20   \\
$\Sigma^+$ & 2.45$\pm$0.01  &  2.61&  2.20 & 2.31  &  2.42 &2.54  &  
2.36 & 2.43     \\

$\Xi^o$    & -1.25$\pm$0.014 &  -1.41 &  -1.10 & -1.16 &  -1.26 & -1.32 &   
-1.20&  -1.24    \\
$\Xi^-$    & -0.65$\pm$0.002 &  -0.50 &  -0.48 & -0.50&  -0.56 & -0.59&
 -0.54 &   -0.56  \\
$\Lambda$ & -0.61$\pm$0.004  &  -0.60 &  -0.54 &   -0.57 & -0.63 & -0.64 &  
-0.57 & -0.59 \\ 
$\Sigma \Lambda$ & 1.61$\pm$0.08 &1.51 &1.53 & 1.50 & 1.77& 1.75& 
1.61 & 1.63 \\ \hline
$\Delta$CG  & 0.49 $\pm$ 0.05 & 0 & 0.29 & 0.31 &0.25  & 0.31 & 
0.44 &  0.48  \\

\end{tabular}
\end{center}
\caption{Comparison of the results of $\chi$QM with the generalized 
Cheng-Li mechanism,  configuration mixing and ``mass adjustments'' 
in units of $\mu_N$ for different sets of quark masses.} \label{masses}
\end{table}

\begin{table}
\begin{center}
\begin{tabular}{ccccccc}     
Quantity &Expt & NRQM & \multicolumn{2}{c}{Without gluon
polarization} & \multicolumn{2}{c}{With gluon polarization} \\ \cline{4-7} 
 & value & & $\chi$QM &$\chi$QM$_{gcm}$ &  $\chi$QM  &
$\chi$QM$_{gcm}$ \\ \hline
$\Delta u$ & 0.85$\pm$0.05 \cite{adams} & 1.33 & 1.02 & 0.95& 0.91& 0.84 \\
$\Delta d$ & -0.41$\pm$0.05 \cite{adams}&-0.33 & -0.38 & -0.31& -0.49& -0.42 \\
$\Delta s$ & -0.07$\pm$0.05 \cite{adams}  & 0 & -0.02& -0.02& -0.13& -0.13 \\
$\Delta \Sigma$ & 0.30$\pm$0.06 \cite{abe}& 1 & 0.62 & 0.62& 0.29& 0.29 \\
$\mu_p$  & 2.79$\pm$0.00 \cite{PDG}  &  2.72 & 3.03 & 2.80 & 3.00 & 2.77 \\
$\mu_n$  & -1.91$\pm$0.00\cite{PDG} &  -1.81 & -2.24  & -1.99 & -2.21& -1.96 \\
$\mu_{\Sigma^-}$ & -1.16$\pm$0.025\cite{PDG} &  -1.01  & -1.26 & -1.20 & -1.23 & -1.17 \\
$\mu_{\Sigma^+}$ & 2.45$\pm$0.01\cite{PDG}  &  2.61& 2.62  & 2.43& 2.59& 2.40 \\
$\mu_{\Xi^o}$    & -1.25$\pm$0.014\cite{PDG} & -1.41 & -1.47 & -1.24 & -1.50 & -1.27 \\
$\mu_{\Xi^-}$    & -0.65$\pm$0.002\cite{PDG} & -0.50 & -0.54  & -0.56 & -0.57 & -0.59 \\
$\mu_{\Lambda}$ & -0.61$\pm$0.004\cite{PDG} & -0.59 & -0.68  & -0.59 & -0.71 & -0.62 \\ 
$\mu_{\Sigma \Lambda}$ & 1.61$\pm$0.08\cite{PDG} & 1.51 & 1.72 & 1.63 & 1.69 & 1.60 \\
\end{tabular}
\end{center} 
\caption{The phenomenological quantities affected by the inclusion of 
gluon polarization. 
The magnetic moments are in units of $\mu_N$.} 
\label{gluon}
\end{table}


\begin{references}
\bibitem{CGSR} The Coleman Glashow sum rule  for 
the baryon magnetic moments is given as 
$\Delta {\rm CG} = 
\mu_p-\mu_n+\mu_{\Sigma^-}-\mu_{\Sigma^+}+\mu_{\Xi^o}-\mu_{\Xi^-}=0.$
Experimentally
$\Delta{\rm CG}=0.49 \pm 0.05$,
clearly depicting the violation of CGSR by ten standard deviations.
\end{references}
\end{document}